\documentclass[%
 reprint,
superscriptaddress,
 amsmath,amssymb,
 aps,
 prl,
]{revtex4-2}

\usepackage{graphicx} 
\usepackage{amsmath}
\usepackage{comment}
\usepackage[scr=rsfso]{mathalfa}

\DeclareMathOperator{\bin}{\hat{\textit{b}}_{\text{in}}}
\DeclareMathOperator{\bout}{\hat{\textit{b}}_{\text{out}}}
\DeclareMathOperator{\dbin}{\hat{\textit{b}}_{\text{in}}^\dagger}
\DeclareMathOperator{\dbout}{\hat{\textit{b}}_{\text{out}}^\dagger}

\usepackage{ifthen}
\usepackage{amsmath}
\usepackage{amssymb}
\usepackage{amsfonts}
\usepackage{amsthm}
\usepackage{physics}
\usepackage{cancel}
\usepackage{mathbbol}
\usepackage{empheq}
\usepackage{xfrac}
\usepackage{latexsym}
\usepackage{cancel}
\usepackage{xcolor}
\usepackage{tcolorbox}
\usepackage{hyperref}
\usepackage{enumitem}
\usepackage{graphicx}
\usepackage{subcaption}
\usepackage{tikz}
\usepackage[normalem]{ulem}

\usepackage{pgfplots}
\pgfplotsset{compat=1.18, width = 12cm, height = 8cm}
\usepackage{pstool}
\usepackage{here}
\usepackage{svg}
\usepackage{overpic}

\usepackage{hyperref}

\DeclareMathOperator{\comma}{~,}
\DeclareMathOperator{\fullstop}{~.}

\usetikzlibrary{decorations.pathmorphing}
\usetikzlibrary{arrows,decorations.pathmorphing}

\usepackage{ragged2e}
\usepackage{physics}
\usepackage{braket}
\usepackage{graphicx}
\usepackage{dcolumn}
\usepackage{bm}

\begin{document}

\preprint{APS/123-QED}

\title{Thermodynamic Framework for Coherently Driven Systems}
\author{Max Schrauwen}
\thanks{These two authors contributed equally}
\affiliation{Department of Physics, RWTH Aachen University, 52056 Aachen, Germany}
\author{Aaron Daniel}
\thanks{These two authors contributed equally}
\author{Marcelo Janovitch}
\author{Patrick P. Potts}
\affiliation{
Department of Physics and Swiss Nanoscience Institute, University of Basel, Klingelbergstrasse 82, CH-4056 Basel, Switzerland\
}
\date{\today}

\date{\today}

\begin{abstract}
The laws of thermodynamics are a cornerstone of physics. At the nanoscale, where fluctuations and quantum effects matter, there is no unique thermodynamic framework because thermodynamic quantities such as heat and work depend on the accessibility of the degrees of freedom. We derive a thermodynamic framework for coherently driven systems, where the output light is assumed to be accessible. The resulting second law of thermodynamics is strictly tighter than the conventional one and it demands the output light to be more noisy than the input light. We illustrate our framework across several well-established models and we show how the three-level maser can be understood as an engine that reduces the noise of a coherent drive. Our framework opens a new avenue for investigating the noise properties of driven-dissipative quantum systems.
\end{abstract}

\maketitle


\textit{\label{sec:level1}Introduction\protect} ---
In the last decades the laws of thermodynamics, traditionally used to describe macroscopic systems, have been extended to small, stochastic systems and even to the quantum regime with great success~\cite{seifert_2012,vandebroek_2015,brandao_2015,Vinjanampathy_2016,Binder2018,deffner_book,strasberg_book,potts2024quantumthermodynamics}. A major obstacle in this endeavor turned out to be the definition of one of the most central quantities: work~\cite{Talkner2007,Perarnau2017,kerremans_2022,elouard_2023}. At the macroscale, work is carried by the accessible, macroscopic degrees of freedom, while heat is carried by the microscopic, inaccessible degrees of freedom. At the nanoscale, all degrees of freedom are microscopic, resulting in ambiguity when defining work. To remedy this situation, definitions for work that depend on the considered task have been suggested~\cite{Niedenzu2019,Gallego_2016,PhysRevLett.124.130601}.

\begin{figure}
    \centering
    \includegraphics[width=.85\linewidth]{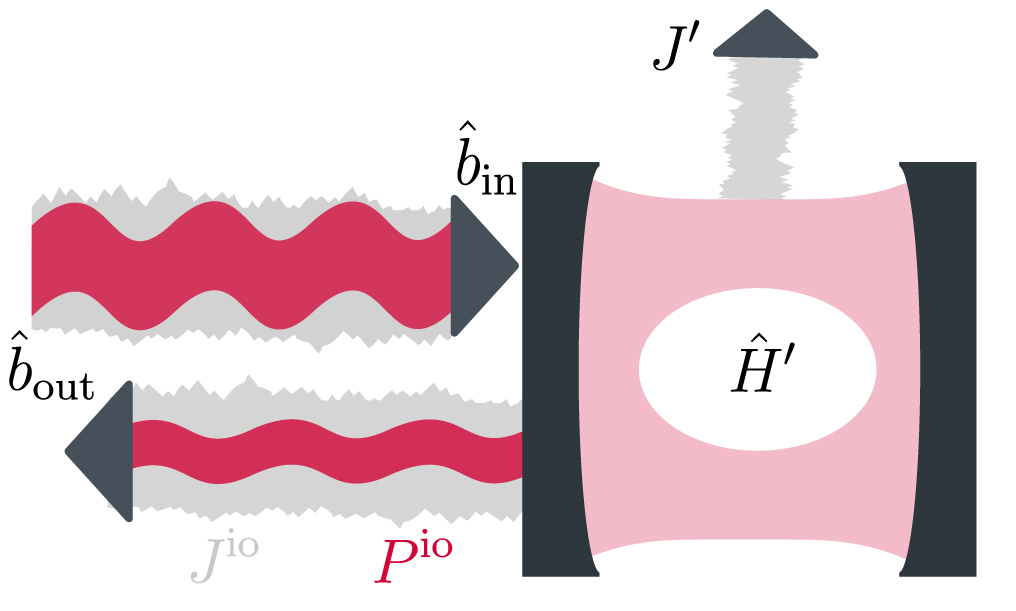}
\caption{\justifying Setup. A cavity with an embedded quantum system, {governed by the Hamiltonian $\hat{H}'$,} is driven by the input field, which contains a coherent part, given by the average of the input field $\langle\hat{b}_{\rm in}\rangle$, as well as thermal fluctuations. Compared to the input, the output field, $\bout$, typically has a smaller coherent part and larger fluctuations. Work ($P^{\rm io}$) and heat ($J^{\rm io}$) are given by the changes in the coherent part and the fluctuations respectively. The heat dissipated by the intracavity system is given by $J'$.}
\label{fig:system}
\end{figure}

The most prominent definition for work (henceforth called the \textit{conventional} definition) relies on a semiclassical description in terms of a time-dependent Hamiltonian. Work is then described by the energy changes due to its time-dependence, which are induced by controllable degrees of freedom that are described classically~\cite{Vinjanampathy_2016,strasberg_book, potts2024quantumthermodynamics}. While often adequate, this approach has conceptual shortcomings for coherently driven systems. To illustrate this, we consider an optical cavity that is driven by coherent light as sketched in Fig.~\ref{fig:system}. In this scenario, the drive can be described by a time-dependent Hamiltonian and the conventional definition can be applied. However, all the energy that leaks out of the cavity is then treated as heat, even if it contains coherence. This may be appropriate if the light leaving the cavity is dissipated but as pointed out in Ref.~\cite{PhysRevLett.124.130601}, this light may be further used. This becomes particularly obvious in the case of an empty cavity, where the light is simply reflected with a shifted phase. The output light leaving the cavity is then just as useful as the input light that induces the time-dependence of the Hamiltonian.

In Ref.~\cite{PhysRevLett.124.130601}, a resolution of this issue was suggested by considering the coherent part (i.e., the displacement) of the output light as work, not heat. 
However, it remains unclear if this intuitive definition for work can be embedded in a thermodynamically consistent framework where a non-negative entropy production is associated to the corresponding definition of heat.

In this Letter, we solve this critical problem by formulating the first and second law of thermodynamics consistent with these new definitions. In this framework, the second law demands that the noise in the output light is larger than the noise in the input light, as long as there is no other source of entropy production. We further show that our new second law is always tighter than the conventional second law. This is consistent with treating the output light as accessible: having more access generally reduces irreversibility.
We illustrate our thermodynamic framework with the help of a Kerr oscillator, a two-level system embedded in a cavity, and a three-level maser~\cite{Niedenzu2019, Li2017}. The last acts as a heat engine, which in our framework corresponds to a machine that can reduce the noise in the light field.

\textit{{Setup} and model} --- We consider a quantum system placed inside a cavity that is subjected to a coherent drive and couples to a bosonic bath via one channel (see Fig.~\ref{fig:system}). The Hamiltonian of our {setup} reads
\begin{align}
    \hat{H}(t) = \Omega \hat{a}^\dagger \hat{a} + \hat{H'} + \hat{H}_{\rm d}(t)~,  \label{eq:Hamiltonian}
\end{align}
where $\Omega$ is the cavity frequency - and  $\hat{H}'$ describes the system inside the cavity as well as its coupling to the cavity mode. {Throughout, primed quantities act on the intracavity system.} The drive term is given by
\begin{align}
    \hat{H}_{\rm d}(t) &= i\sqrt{\kappa} \left(\langle \dbin(t)\rangle\hat{a} -\langle\bin(t)\rangle\hat{a}^\dagger \right) \comma 
    \label{eq:drive}
\end{align}
with $\kappa$ denoting the linewidth of the cavity.
We further introduced the input field which describes the coherent drive as well as the thermal noise entering the cavity~\cite{gardiner_1985}
\begin{equation}
\label{eq:drive}
    \langle\bin(t)\rangle = f(t) e^{-i\omega_{\rm d}  t} ,\hspace{.25cm} \langle\!\langle\dbin(t)\bin(t')\rangle\!\rangle = n_{\rm c}\delta(t-t'),
\end{equation}
where $n_{\rm c}$ denotes the thermal occupation and $\langle\!\langle \hat{A}\hat{B}\rangle\!\rangle = \langle \hat{A}\hat{B}\rangle-\langle\hat{A}\rangle\langle\hat{B}\rangle$. We consider a drive at frequency $\omega_{\rm d}$ with a potential slow modulation, i.e., $|\partial_tf(t)|/|f(t)|\ll\omega_{\rm d}$. We note that a coherent drive assumes a source with a stable phase. Any phase can then be defined with respect to the source, which provides a phase reference.

We model the dynamics of our setup with the quantum master equation,
\begin{equation}
    \partial_t \hat{\rho}(t) = -i \qty[\hat{H}(t),\hat{\rho}(t)] + \mathcal{L}_{\rm c} \hat{\rho}(t) + \mathcal{L}' \hat{\rho}(t)\comma\label{eq:system_QME}
\end{equation}
where $\mathcal{L}'$ is an arbitrary Lindbladian dissipator acting on the intracavity system and the dissipator of the cavity mode reads 
\begin{align}
\label{eq:dissc}
    \mathcal{L}_{\rm c} \hat{\rho} &= \kappa n_{\rm c} \mathcal{D}[\hat{a}^\dagger] \hat{\rho} + \kappa (n_{\rm c}+1) \mathcal{D}[\hat{a}]\hat{\rho}  \comma 
\end{align}
with $\mathcal{D}[\hat{a}]\hat{\rho} = \hat{a} \hat{\rho} \hat{a}^\dag - \frac{1}{2}\{ \hat{a}^\dag\hat{a},\hat{\rho}\}$.
We assume that photons can only be exchanged with the input channel and, thus, demand that $\Tr{\hat{a}^\dagger\hat{a}\mathcal{L}'\hat{\rho}}=0$. Additional dissipation channels for the photons can be included (see Supplemental Material~\cite{supp}).
To describe the light that is reflected by the cavity, we introduce the output operator~\cite{gardiner_1985},
\begin{align}
    \hat{b}_\text{out} = \hat{b}_\text{in} + \sqrt{\kappa}\hat{a} \fullstop \label{eq:inout}
\end{align}

\textit{Thermodynamics} ---
\label{sec:level2}  
The first law of thermodynamics states that any change in the internal energy $U$ of a system can be divided into heat and work
\begin{equation}
\label{eq:firstlaw}
    \partial_t U = P + J \comma
\end{equation}
where $P$ denotes power (work per unit time) and $J$ the heat current. The second law constrains the physically allowed processes by demanding a non-negative entropy production defined as 
\begin{align}
    \dot{\Sigma}\equiv k_{\rm B}\partial_tS_{\rm vN}(\hat{\rho})-\frac{J_{\rm c}}{T_{\rm c}}-\frac{J'}{T'} \geq 0 \comma \label{eq:secondlaw}
\end{align}
with the von Neumann entropy $S_{\text{vN}}(\hat{\rho}) = - \Tr{\hat{\rho} \ln \hat{\rho}}$. In Eq.~\eqref{eq:secondlaw}, the heat current is split into two parts, $J=J_{\rm c}+J'$, corresponding to the dissipation of the cavity and the intracavity system, respectively.  Here we assume one temperature for the environment of the cavity and one for the environment of the intracavity system. Additional temperatures are discussed in  Supplemental Material~\cite{supp}.

Before introducing our new thermodynamic framework, we recapitulate how thermodynamic quantities can be consistently defined in the conventional framework, where energy changes due to the time dependence of the Hamiltonian are interpreted as work, i.e., $P=\langle\partial_t\hat{H}(t)\rangle$. Typically, the internal energy is defined as the expectation value of $\hat{H}(t)$, and heat is defined through the first law in Eq.~\eqref{eq:firstlaw} \cite{Alicki_1979,kosloff_1984, Vinjanampathy_2016}. This approach may result in violations of the second law for master equations, which do not rely on the secular approximation \cite{Novotny_2002,Levy_2014,Trushechkin_2016,Hofer_2017,Gonzalez_2017,Potts_2021}, as is the case in our Eq.~\eqref{eq:system_QME}. Here we follow the approach in Ref.~\cite{Potts_2021}, which ensures thermodynamic consistency. To this end, we introduce the thermodynamic Hamiltonian, 
\begin{align}
\hat{H}_{\rm TD} = \omega_{\rm d} \hat{a}^\dagger \hat{a} + \hat{H}_{\rm TD}'  \comma \label{eq:TD_hamiltonian}
\end{align}
which is used to define the internal energy as $U=\langle \hat{H}_{\rm TD}\rangle$.  Note that $\hat{H}_{\rm TD}$ may differ from the Hamiltonian $\hat{H}(t)$ that governs the dynamics~\cite{supp}.
 This choice reflects the fact that the drive populates the cavity with photons at $\omega_{\rm d}$. Thermodynamic consistency is ensured by choosing $n_{\rm c} =[e^{\omega_{\rm d}/(k_{\rm B}T_{\rm c})}-1]^{-1}$ in Eq.~\eqref{eq:dissc}. We note that our master equation is only valid when the frequency dependence of the Bose-Einstein distribution within the cavity linewidth can be neglected. The term $\hat{H}'_{\rm TD}$ acts on the intracavity system, quantifying its internal energy. A more detailed discussion on the thermodynamic Hamiltonian can be found in Refs.~\cite{supp,Potts_2021}.

With the thermodynamic Hamiltonian, we find the first law of thermodynamics with
\begin{equation}
    P = -i\langle [\hat{H}_{\rm TD},\hat{H}(t)]\rangle,\hspace{.5cm} J =  \Tr{
    \hat{H}_{\rm TD}\mathcal{L}\hat{\rho}(t)},
    \label{eq:powerandheat}
\end{equation}
where $\mathcal{L}=\mathcal{L}_{\rm c}+\mathcal{L}'$.
We recover the standard form of power when the thermodynamic Hamiltonian obeys $-i [\hat{H}_{\rm TD}, \hat{H}(t)] \simeq \partial_t \hat{H}(t)$.
In our case this holds as long as $-i[\hat{H}_{\rm TD},\hat{H}']=\partial_t\hat{H}'$~\cite{supp}. For simplicity, we consider a time-independent $\hat{H}'$ from here on, such that the power is associated only to $\hat{H}_{\rm d}$.

As above, the heat current may be decomposed in two terms $J=J_{\rm c}+J'$, with~\cite{supp}
\begin{align}
\label{eq:heatdecomp}
     J_{\rm c} &= \Tr{\omega_{\rm d} \hat{a}^\dagger \hat{a}
    \mathcal{L}_{\rm c}\hat{\rho}(t)}\comma &  J' &= \Tr{
    \hat{H}_{\rm TD}'\mathcal{L}'\hat{\rho}(t)}  \fullstop
\end{align}
For the heat current associated to the cavity and the power, we can evaluate these expressions to find 
\begin{align}
    J_{\rm c} &= \omega_{\rm d}\kappa (n_{\rm c} - \langle \hat{a}^\dagger \hat{a} \rangle) \comma \label{eq:heat_conventional_cavity}\\
    P &= -\sqrt{\kappa}\omega_{\rm d} \bigg( \langle \hat{b}_{\text{in}}^\dagger(t) \rangle \langle \hat{a} \rangle + \langle \hat{b}_{\text{in}}(t) \rangle \langle \hat{a}^\dagger  \rangle \bigg) \fullstop\label{eq:power_conventional_cavity}
\end{align}
Typically, the heat characterizes the amount of energy exchanged with the bath, while the power quantifies the energy exchanged with the coherent driving field. This interpretation is challenged in the present scenario, as both the coherent drive and the thermal noise are carried by the same environment. Indeed, one may show that the total energy exchanged with the cavity environment is equal to~\cite{supp}
\begin{equation}
    \label{eq:toten}
    \omega_{\rm d}\left(\langle\dbout(t)\bout(t)\rangle-\langle\dbin(t)\bin(t)\rangle\right) = P+J_{\rm c}.
\end{equation}

With the definition of heat above, it can be shown by using Spohn's inequality \cite{spohn1978,spohnlebowitz1978} that the second law in Eq.~\eqref{eq:secondlaw} is satisfied if $\mathcal{L}'e^{-\hat{H}_{\rm TD}'/(k_{\rm B}T')}=0$~\cite{supp}.

\textit{New thermodynamic framework} ---
We put forward new definitions of heat and work by decomposing the energy flow in Eq.~\eqref{eq:toten} into the coherent part of the fields and their variances,
\begin{align}
     P^{\rm io} &=- \omega_{\rm d}\left(|\langle\bout(t)\rangle|^2 -|\langle\bin(t)\rangle|^2\right) \comma \label{eq:Pio}
    \\J_{\rm c}^{\rm io} &= -\omega_{\rm d}\left(\langle\!\langle\dbout(t)\bout(t)\rangle\!\rangle -\langle\!\langle\dbin(t)\bin(t)\rangle\!\rangle\right). \label{eq:Jio}
\end{align}
In these definitions, the power is determined by the coherent part of the input and output light. Positive power implies that the input light carries more coherence than the output light. Similarly, a negative heat current implies that the output light is more noisy than the input light. Equation \eqref{eq:Pio} has been suggested previously as a definition for work \cite{PhysRevLett.124.130601} since the coherent part of the output light can be further used to drive another quantum system. 
Equations \eqref{eq:Pio} and \eqref{eq:Jio} can, thus, be understood as the relevant definitions of heat and work if the output light is accessible, while Eqs.~\eqref{eq:heat_conventional_cavity} and \eqref{eq:power_conventional_cavity} are the relevant definitions if all the light leaving the cavity is dissipated. We note that the same definitions have been found in a qubit coupled to a waveguide using bipartite frameworks~\cite{2021_maffei,2024_prasad}. For an alternative thermodynamic treatment of coherently driven two-level systems, see Ref.~\cite{2025_soret}.

By definition we find $P^{\rm io} + J_{\rm c}^{\rm io}  = P +J_{\rm c} $, guaranteeing that the first law holds for the new definitions.
Here we prove the second law and elevate Eqs.~\eqref{eq:Pio} and \eqref{eq:Jio} to a consistent thermodynamic framework. To this end, we exploit the fact that the master equation in Eq.~\eqref{eq:system_QME} can be written in a different form by letting $\mathcal{L}_{\rm c}\rightarrow\mathcal{L}_{\rm s}$ and $\hat{H}\rightarrow \hat{H}_{\rm s}$ with
\begin{align}
    \mathcal{L}_{\rm s} &= \kappa n_{\rm c} \mathcal{D}[\hat{a}^\dagger - \langle \hat{a}^\dagger \rangle] 
    + \kappa (n_{\rm c}+1) \mathcal{D}[\hat{a} - \langle \hat{a} \rangle] \comma\\
    \hat{H}_{\rm s} &= \hat{H} +\frac{i}{2}\kappa\left(\langle \hat{a}^\dagger\rangle \hat{a}  - \langle \hat{a}\rangle \hat{a}^\dagger \right)  \fullstop \label{eq:shifted_Hamiltonian} 
\end{align}
Importantly, this transformation leaves the master equation invariant \cite{landi_2024}.
Heat and work may then be computed in full analogy to Eq.~\eqref{eq:powerandheat}, and we find
\begin{align}
    P^{\rm io}& = -i\langle [\hat{H}_{\rm TD},\hat{H}_{\rm s}(t)]\rangle,\\
    J_{\rm c}^{\rm io} &=  \Tr{\omega_{\rm d}\hat{a}^\dagger\hat{a}\mathcal{L}_{\rm s}\hat{\rho}(t)}=\omega_{\rm d}\kappa (n_{\rm c} - \langle\!\langle \hat{a}^\dagger \hat{a} \rangle\!\rangle) .
    \label{eq:powerandheat2}
\end{align}
If $\partial_t \langle \hat{a}\rangle \simeq -i\omega_{\rm d} \langle \hat{a}\rangle$,
we find that the power can again be obtained from the standard expression $P^{\rm io} = \langle \partial_t \hat{H}_{\rm s}(t) \rangle$. 
We note that the rewriting of the master equation only has the effect $P\rightarrow P^{\rm io}$ and $J_{\rm c}\rightarrow J_{\rm c}^{\rm io}$ and leaves $J'$ invariant. 

The second law again follows from Spohn's inequality~\cite{supp} and we find
\begin{align}
    \dot{\Sigma}^{\rm io}\equiv k_{\rm B}\partial_tS_{\rm vN}(\hat{\rho})-\frac{J_{\rm c}^{\rm io}}{T_{\rm c}}-\frac{J'}{T'} \geq 0 \fullstop \label{eq:secondlaw2}
\end{align}
Furthermore, we find that the second law is strictly tighter than in the conventional framework,  
\begin{equation}
    \dot{\Sigma}  \geq \dot{\Sigma}^{\rm io}  \geq 0 \fullstop \label{eq:entropy_ineq}
\end{equation}
With this we have shown that the new definitions for heat and work are thermodynamically consistent. 

We note that the nonunique splitting of master equations into a unitary and a dissipative part has been leveraged before to describe the thermodynamics of systems coupled strongly to their environment~\cite{colla_2022}.

\textit{Empty cavity} ---
As a first example to illustrate our new thermodynamic framework, we consider the case where there is no additional system inside the cavity, i.e., $\hat{H}' = \hat{H}_{\rm TD}'=0$, and there is no additional loss channel, $\mathcal{L}'=0$. 
In our new framework we find that, contrary to the standard definitions, the power and heat current vanish in steady state, $P^{\rm io}=J^{\rm io}=0.$ 
This reflects the fact that the input is simply reflected by the cavity and, up to a phase, the output light is identical to it. The output light is, thus, just as valuable as the input light.
\begin{figure*}
    \centering
    \includegraphics[width = \linewidth]{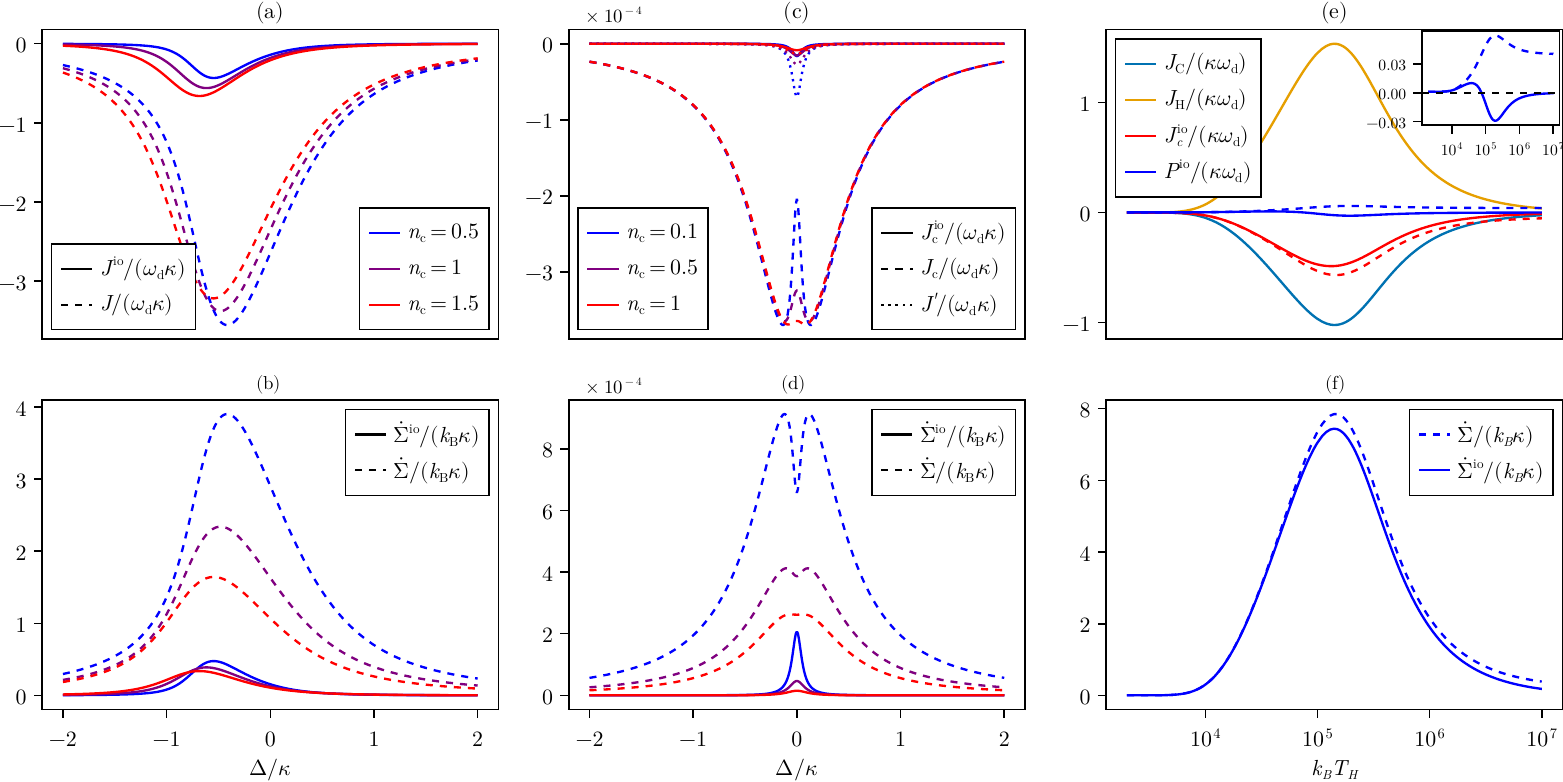}
    \caption{ \justifying Conventional and new definitions of heat and work. For (a)-(d) $\Omega/\kappa =10^4$, $\gamma/\kappa = 0.05 $, $g/\kappa =0.1$, $f/\kappa =0.01 $.  (a) Heat current and (b) entropy production for the Kerr oscillator for different occupation numbers $n_{\rm c}$ as a function of detuning $\Delta = \Omega -\omega_{\rm d}$. Note that  $P^{({\rm io})} = - J^{({\rm io})}$. Parameters: $\Omega/\kappa =10^4 $, $K/\kappa = 0.05 $, $f/\kappa =1 $. (c) Heat currents and (d) entropy production for the two-level system for different occupation numbers as a function of detuning. (e)-(f) Heat currents and power, and entropy production for the quantum three-level maser; $\Omega/\kappa = 10^4,~g/\kappa=0.7,~f/\kappa=0.1,~\omega_2/\kappa=\omega_3/(3\kappa)=\Omega/\kappa,~\gamma_H/\kappa=0.5,~\gamma_C/\kappa = 100,~n_c=4.54,~n_C=0.007$. (e) We observe that there is a region, $k_BT_H\approx10^5$ in which the system is a heat engine, $P^\text{io}<0$, while $P>0$ (see also inset). The dashed black line in the inset indicates the zero. (f) Entropy production peaks at the heat engine regime.}
    \label{fig:panel}
\end{figure*}

\textit{Kerr oscillator} --- 
As a second, nonlinear example we consider a Kerr oscillator described by $\hat{H}' = K \hat{a}^\dagger \hat{a}^\dagger \hat{a}\hat{a},$
with Kerr parameter $K$. 
We do not consider additional losses $\mathcal{L}'=0$, and we consider the thermodynamic Hamiltonian $\hat{H}_{\rm TD}=\omega_{\rm d}\hat{a}^\dagger\hat{a}$, which assigns each photon within the cavity the energy $\omega_{\rm d}$. We note that this requires $K\ll\omega_{\rm d}$ such that a constant thermal occupation $n_{\rm c}$ is justified even though the cavity frequency effectively depends on the number of photons within the cavity. 
We compare the conventional definitions of heat and work with our new definitions (see Fig. \ref{fig:panel}). As the temperature is increased, the conventional definition predicts a decrease in entropy production because fewer photons can enter the cavity from the drive due to the photon-number dependence of the resonance frequency. In contrast, our new definition is much less sensitive to an increase in $n_{\rm c}$ since the entropy production only arises from a reduced coherence due to nonlinear effects. This illustrates how the two definitions provide different insight into irreversibility.

\textit{Two-level system} ---
As a next example, we consider a two-level system embedded in the cavity described by
\begin{align}
    \hat{H}' &= \frac{\Omega}{2} \hat{\sigma}_{\rm z} + g (\hat{a}^\dagger \hat{\sigma}_- + \hat{a} \hat{\sigma}_+) \comma\\ 
    \hat{H}_{\rm TD} &= \omega_{\rm d} \left(\hat{a}^\dagger\hat{a}+\hat{\sigma}_+\hat{\sigma}_-\right) \comma \\ 
    \mathcal{L}' &= \gamma n_{\rm q} \mathcal{D}[\hat{\sigma}_+]  + \gamma (n_{\rm q}+1) \mathcal{D}[\hat{\sigma}_-] \comma
\end{align}
where $g$ is the coupling between the cavity and the qubit, $\gamma$ and $n_{\rm q}=[e^{\omega_{\rm d}/(k_{\rm B}T_{\rm q})}-1]^{-1}$ denote the coupling and the occupation of an additional thermal reservoir coupled to the qubit, and the system operators of the two-level system are given by  $
\hat{\sigma}_{\rm z} = \dyad{e} - \dyad{g}\comma ~ \hat{\sigma}_-  = \dyad{g}{e} = \hat{\sigma}_+^\dagger$. The thermodynamic Hamiltonian assigns each excitation the same energy $\omega_{\rm d}$.

We find that the additional heat current for the intracavity system can be expressed as follows:
\begin{align}
    J' =   \omega_{\rm d} \gamma \frac{n_{\rm q}}{n_{\rm F}}(n_{\rm F} - \langle \hat{\sigma}_+\hat{\sigma}_- \rangle) \comma
\end{align}
where $n_{\rm F} =[e^{\omega_{\rm d}/(k_{\rm B}T_{\rm q})}+1]^{-1}$ denotes the Fermi-Dirac occupation. Here we focus on a single-temperature environment $n_{\rm c} = n_{\rm q}$.

For a weak drive on resonance ($\Delta=0$), the light emitted by the two-level system destructively interferes with the drive, preventing the cavity from being filled \cite{auffeves_2007}. As a consequence, the conventional entropy production is reduced. In contrast, our new definition shows a peak in the entropy production at $\Delta = 0$ because part of the light is dissipated through the two-level system. Increasing temperature suppresses coherence in the two-level system. As a consequence, the features resulting from destructive interference are suppressed and we find a finite $J^{\rm io}$ as the coherence of the light is reduced.

\textit{Three-level maser}---
The three-level maser is a canonic example of a quantum heat engine~\cite{Geusic1967, kosloff_1984, Li2017, Dorfman2018, Niedenzu2019,Mitchison2019, Kalaee_2021}. It can be described by
\begin{align} 
H' &= \omega_3\dyad{3} + \omega_2\dyad{2} + g(\hat{a}\dyad{2}{1} + \hat{a}^\dagger\dyad{1}{2}),\\
H_\text{TD} &= \omega_{\rm d}(\hat{a}^\dagger\hat{a} + \dyad{2}) + \omega_3\dyad{3}, 
\end{align} 
with $\omega_3 > \omega_2$. The thermodynamic Hamiltonian extends the two-level case by adding the state $\ket{3}$. Dissipation is described by $\mathcal{L}' = \mathcal{L}_\text{H} + \mathcal{L}_\text{C}$, with 
\begin{align} 
\mathcal{L}_\text{H} &= \gamma_H n_\text{H}\mathcal{D}[\dyad{3}{1}] + \gamma_H (n_{\rm H}+1)\mathcal{D}[\dyad{1}{3}],\\ \mathcal{L}_\text{C} &= \gamma_C n_\text{C}\mathcal{D}[\dyad{3}{2}] + \gamma_C (n_{\rm C}+1)\mathcal{D}[\dyad{2}{3}], \end{align} 
where $n_{\rm H} = [e^{\omega_3/(k_{\rm B}T_{\rm H})}-1]^{-1}$, $n_{\rm C} = [e^{(\omega_3-\omega_{\rm d})/(k_{\rm B}T_{\rm C})}-1]^{-1}$, and $T_{\rm H}>T_{\rm C}$ creates a population inversion between the states $\ket{1}$ and $\ket{2}$.
Due to the quantum mechanical treatment of the cavity field, the conventional definitions of heat and work predict $P>0$ and suggest that the system ceases to operate as a heat engine, see Fig.\,\ref{fig:panel}~(e)~\cite{Li2017,Niedenzu2019}. The reason for this is the energetic cost of maintaining coherence in the cavity.
Our new framework resolves this flaw, as there is no energetic cost to maintain coherence when the output light can be further used, identifying regions in which $P_\text{io}<0$. In these regions, a heat current between hot and cold baths increases the coherence in the output light.

\textit{Entropy production as loss of information} --- Before concluding, we provide an alternative derivation of our main results, connecting to a broader concept of entropy production that goes beyond Markovian master equations. Following Ref.~\cite{Esposito_2010}, Eq.~\eqref{eq:secondlaw} may be obtained from
\begin{equation}
\label{eq:srel}
    \Sigma = k_B S[\hat{\rho}_{\rm tot}||\hat{\rho}\otimes \hat{\tau}],
\end{equation}
where $S[\hat{\rho}||\hat{\sigma}]={\rm Tr}\{\hat{\rho}\ln\hat{\rho}-\hat{\rho}\ln\hat{\sigma}\}$, $\hat{\rho}_{\rm tot}$ describes the density matrix of the system and environment, $\hat{\rho} = {\rm Tr}_{\rm B}\{\hat{\rho}_{\rm tot}\}$ is obtained by tracing out the environment, and $\hat{\tau}$ denotes the Gibbs state of the environment. In Eq.~\eqref{eq:srel}, the entropy can be understood as the information that is lost when describing the environment using only temperature~\cite{potts2024quantumthermodynamics}. For coherently driven systems, the environment also contains the coherent drive, as well as the response of the system. An appropriate state to describe the environment is then a displaced thermal state. We may obtain our new second law through 
\begin{equation}
\label{eq:srelio}
    \Sigma^{\rm io} = k_B S[\hat{\rho}_{\rm tot}||\hat{\rho}\otimes \hat{\tau}({\mathbf \alpha})],
\end{equation}
where $\hat{\tau}({\bf \alpha})$ denotes a displaced thermal state, with each bosonic bath mode $\hat{b}_k$ displaced by its average $\alpha_k={\rm Tr}\{\hat{b}_k\hat{\rho}_{\rm tot}\}$~\cite{supp}. Equation \eqref{eq:srelio} may serve as a starting point to extend our framework to strong coupling scenarios, where dissipation is no longer described by Eq.~\eqref{eq:dissc}~\cite{Beaudoin_2011,mercurio_2022}, or to include further access on the output light, such as its squeezing.

\textit{Conclusions and outlook} --- 
We have shown how treating the output light of a coherently driven system as accessible leads to a consistent thermodynamic framework. This framework features a second law that is strictly tighter than the conventional approach. Without an additional entropy source, our second law implies that the noise in the output light is always larger than the noise in the input light.
Our framework is, thus, particularly promising for investigating the noise properties of cascaded systems and opens up novel ways of investigating the thermodynamics of optical devices such as circulators and amplifiers. Of particular interest is the thermodynamics of nonreciprocal devices \cite{Metelmann_2015}.

Our formalism only treats the coherent part of the output light as accessible. This requires the presence of a phase reference. Quantifying work in the output light in the absence of a perfect phase reference remains an open problem. Another promising avenue is to investigate fluctuations of heat and work in the new framework, which would allow for investigating fluctuation theorems \cite{seifert_2012} and thermodynamic uncertainty relations \cite{Horowitz_2020}.

\begin{acknowledgments}
\textit{Acknowledgment} --- We acknowledge fruitful discussions with Juliette Monsel, Ariane Soret, and Mario Berta.  This work was supported by the Swiss National Science Foundation (Eccellenza Professorial Fellowship PCEFP2\_194268).
\end{acknowledgments}
\textit{Data availability} --- The data that support the findings of this article are not publicly available. The data are available from the authors upon reasonable request.

\bibliography{ref_final}
\newpage

\widetext
\begin{center}
\vskip0.5cm
{\Large Supplemental Material: A Thermodynamic Framework for Coherently Driven Systems}
\vskip0.2cm
Max Schrauwen,$^1$ Aaron Daniel,$^2$ Marcelo Janovitch,$^2$ Patrick P. Potts$^2$
\vskip0.1cm
\textit{$^1$ Department of Physics, RWTH Aachen University, 52056 Aachen, Germany\\ $^2$ Department of Physics and Swiss Nanoscience Institute,\\
University of Basel, Klingelbergstrasse 82, 4056 Basel, Switzerland}
\vskip0.1cm
\end{center}
\vskip0.4cm

\setcounter{equation}{0}
\setcounter{figure}{0}
\setcounter{table}{0}
\setcounter{page}{1}
\renewcommand{\theequation}{S\arabic{equation}}
\renewcommand{\thefigure}{S\arabic{figure}}

In this Supplemental Material, we provide additional information and calculations to the statements made in the Letter. In Sec.~I, we {discuss the thermodynamic Hamiltonian and show how the power is obtained from it}. Section II illustrates the assumptions that result in the decomposition of heat given in Eq.~(11). In Sec.~III, we provide the power and the heat current of the cavity using input-output operators. Section IV provides proofs for the second law of thermodynamics in both the conventional and the new framework {and Sec.~V provides generalizations to additional dissipation channels and multiple temperatures. Finally, Sec.~VI provides an alternative derivation of our framework, where entropy production is understood as a loss of information.}

\section{I. The Thermodynamic Hamiltonian}

{The thermodynamic Hamiltonian, $\hat{H}_{\rm TD}$, is the operator that quantifies the internal energy in a thermodynamically consistent way for the Markovian master equations considered in the manuscript. Importantly, it does not have to be equal to the Hamiltonian $\hat{H}(t)$ which enters the master equation. As discussed in Ref.~\cite{Potts_2021}, the thermodynamic Hamiltonian obeys
\begin{equation}
    \mathcal{L}_{\rm c} e^{-\beta\hat{H}_{\rm TD}} = 0.
\end{equation}
with [c.f.~Eq.~(5) in the main text]
\begin{equation}
\label{eq:disscapp}
     \mathcal{L}_{\rm c} \hat{\rho} = \kappa n_{\rm c} \mathcal{D}[\hat{a}^\dagger] \hat{\rho} + \kappa (n_{\rm c}+1) \mathcal{D}[\hat{a}]\hat{\rho},
\end{equation}
  this results in the thermodynamic Hamiltonian
\begin{equation}
\label{eq:htdapp}
    \hat{H}_{\rm TD} = \omega_{\rm d} \hat{a}^\dagger \hat{a} + \hat{H}_{\rm TD}',
\end{equation}
with $\hat{H}_{\rm TD}'$ denoting the internal energy of the intra-cavity system, obeying $[\hat{a},\hat{H}_{\rm TD}']=0$. We note that as long as the cavity dissipation can be modeled through Eq.~\eqref{eq:disscapp}, the coupling between the cavity and the intra-cavity system is small enough such that neglecting it in the internal energy is justified.

The thermodynamic Hamiltonian in Eq.~\eqref{eq:htdapp} assigns each photon the frequency $\omega_{\rm d}$, which is the same frequency that appears in the Bose-Einstein occupation
\begin{equation}
    n_{\rm c} \equiv n_{\rm BE}(\omega_{\rm d})=\frac{1}{e^{\omega_{\rm d}/(k_{\rm B}T_{\rm c})}-1}.
\end{equation}
The fact that only a single number $n_{\rm c}$ parametrizes the dissipation in Eq.~\eqref{eq:disscapp} is a consequence of the Markovian assumption. The lifetime broadening of the cavity resonance is neglected and photons are approximated to be exchanged with the environment at a single frequency. Here we choose this frequency to be equal to the drive frequency (not the cavity frequency) because the drive populates the cavity with photons at frequency $\omega_{\rm d}$. We note that we require $\omega_{\rm d}$ to lie within the linewidth of the cavity in order for the drive to have any effect. In this case, the Markovian approximation resulting in the master equation in the main text requires $n_{\rm BE}(\omega_{\rm d})\simeq n_{\rm BE}(\Omega)$. The approximation further requires $\kappa\ll\Omega$. For this reason $\Omega\langle\hat{a}^\dagger\hat{a}\rangle=\omega_{\rm d}\langle\hat{a}^\dagger\hat{a}\rangle+\mathcal{O}(\kappa)$ implying that $\Omega$ could also be used in the thermodynamic Hamiltonian. However, the laws of thermodynamics are only guaranteed when the same frequency is used in the thermodynamic Hamiltonian and in the Bose-Einstein occupation. Furthermore, using $\omega_{\rm d}$ allows for the power to be expressed through the time-dependence of the Hamiltonian, see below. 
This freedom of varying the precise frequency that appears in the thermodynamic Hamiltonian is justified by the Markovian assumption, which neglects the life-time broadening of the cavity frequency. In this approximation, the precise energy resolution of the exchanged photons is lost~\cite{Potts_2021}. We may exploit this lack of resolution to choose a frequency that ensures the laws of thermodynamics.

Very similar considerations hold for the intra-cavity system. The Markovian assumption requires a separation of time-scales, where the energy of excitations (e.g., photons) are much larger than any broadening introduced by couplings. The thermodynamic Hamiltonian neglects the couplings and assigns the excitations the energy required for thermodynamic consistency. The values of these energies typically have to match the arguments in the bath occupation functions that appear in the dissipators. We stress that while the couplings do not appear in the thermodynamic Hamiltonian, and thus the internal energy, they are nevertheless crucial for the dynamics and therefore appear in the master equation.
Due to the separation of energy scales, which is needed for the Markovian master equation to be justified, we have $\langle \hat{H}_{\rm TD}\rangle \simeq \langle\hat{H}(t)\rangle$. However, only the use of the thermodynamic Hamiltonian guarantees that the laws of thermodynamics are exactly satisfied.}

\subsection{{Power in the} conventional framework}
We show that the commutator of the thermodynamic Hamiltonian $\hat{H}_{\rm TD}=\omega_{\rm d}\hat{a}^\dagger\hat{a}+H_{\rm TD}'$ with the full system Hamiltonian $ \hat{H}(t)$ is approximately given by the time derivative of the full system Hamiltonian $\partial_t \hat{H}(t)$
\begin{equation}
\label{eq:commutator}
    [\hat{H}_{\rm TD}, \hat{H}(t)] =  [\omega_{\rm d} \hat{a}^\dagger\hat{a}, \hat{H}_{\rm d}(t)] + [\hat{H}_{\rm TD}, \hat{H}'(t)] + [\hat{H}_{\rm TD}', \Omega \hat{a}^\dagger\hat{a} ] + [\hat{H}_{\rm TD}', \hat{H}_{\rm d}(t)] .
\end{equation}
We now use the assumptions that the thermodynamic Hamiltonian of the intra--cavity system commutes with the cavity mode, i.e., $[\hat{a},\hat{H}_{\rm TD}']=0$. As a consequence
\begin{equation}
   [\hat{H}_{\rm TD}', \hat{H}_{\rm d}(t)] = [\hat{H}_{\rm TD}', \Omega \hat{a}^\dagger\hat{a} ] = 0 \comma 
\end{equation}
we further assume that
\begin{equation}
      [\hat{H}_{\rm TD}, \hat{H}'(t)] \simeq \partial_t\hat{H}'(t) \fullstop
\end{equation}
For the final term of Eq. \eqref{eq:commutator} we find
\begin{equation}
     [\omega_{\rm d} \hat{a}^\dagger\hat{a}, \hat{H}_{\rm d}] =-i\omega_{\rm d}\sqrt{\kappa}\left(\langle\hat{b}_{\rm in}^\dagger(t)\rangle\hat{a}+\langle\hat{b}_{\rm in}(t)\rangle\hat{a}^\dagger\right)\fullstop
\end{equation}
For $\langle\hat{b}_{\rm in}(t)\rangle=f(t)e^{-i\omega_{\rm d}t}$, we find
\begin{equation}
\label{eq:commdr}
     -i[\omega_{\rm d} \hat{a}^\dagger\hat{a}, \hat{H}_{\rm d}(t)] \simeq \partial_t\hat{H}_{\rm d}(t)\comma
\end{equation}
as long as $\omega_{\rm d}|f(t)|\gg|\partial_tf(t)|$ such that we can neglect the time derivative of $f(t)$ in Eq.~\eqref{eq:commdr}.

Together, the above relations provide 
\begin{equation}
    -i [\hat{H}_{\rm TD}, \hat{H}(t)] \simeq \partial_t \hat{H}(t) \fullstop
    \label{eq:tddt}
\end{equation}
 {We note that this only holds when the internal energy assigned to the photons by the thermodynamic Hamiltonian matches the frequency of the drive. If the cavity frequency is used in the thermodynamic Hamiltonian instead, Eq.~\eqref{eq:tddt} is only approximately true.}

\subsection{{Power in the} new framework}
For the new framework, the calculation goes analogously. However, the drive Hamiltonian needs to be replaced with [c.f.~Eq.~(19)]
\begin{equation}
    \hat{H}_{\rm s} = i\frac{\sqrt{\kappa}}{2} \left(\left[\langle \dbin(t)\rangle+\langle\dbout(t)\rangle\right]\hat{a} -\left[\langle\bin(t)\rangle+\langle\bout(t)\rangle\right]\hat{a}^\dagger \right) \fullstop
\end{equation}
As long as $\langle \hat{a}\rangle(t) = \alpha(t)e^{-i\omega_{\rm d}t}$, where $\alpha(t)$ is a slowly varying function such that $ \omega_{\rm d}|\alpha(t)|\gg\partial_t\alpha(t)$, we find again
\begin{equation}
     -i[\omega_{\rm d} \hat{a}^\dagger\hat{a}, \hat{H}_{\rm s}] \simeq \partial_t\hat{H}_{\rm s}(t)\comma
\end{equation}
such that power can still be expressed through the time-derivative of the Hamiltonian.

\section{II. Decomposition of Heat}
The heat current given in Eq.~(10) can be written as as 
\begin{equation}
    J =  \Tr{\hat{H}_{\rm TD}\mathcal{L}\rho(t)} = \Tr{\omega_{\rm d} \hat{a}^\dagger\hat{a}\mathcal{L}_{\rm c}\rho(t)} +
    \Tr{\omega_{\rm d} \hat{a}^\dagger\hat{a}\mathcal{L}'\rho(t)} +
    \Tr{\hat{H}_{\rm TD}'\mathcal{L}_{\rm c}\rho(t)} +
    \Tr{\hat{H}_{\rm TD}'\mathcal{L}'\rho(t)} \fullstop
\end{equation}
From $[\hat{a},\hat{H}_{\rm TD}']=0$ one may show that
\begin{equation}
    \Tr{\hat{H}_{\rm TD}'\mathcal{L}_{\rm c}\rho(t)} = 0 \fullstop
\end{equation}
In addition, we write
\begin{equation}
    \mathcal{L}'\hat{\rho} = \sum_k \mathcal{D}[\hat{L}_k]\hat{\rho},
\end{equation}
and we assume that $[\hat{L}_k,\hat{a}^{(\dagger)}] =0$. It follows that
\begin{equation}
    \Tr{\omega_{\rm d} \hat{a}^\dagger\hat{a}\mathcal{L}'\rho(t)}  = 0 \fullstop
\end{equation}
We thus find the decomposition of heat provided in the main text
\begin{equation}
    J =  \Tr{\omega_{\rm d} \hat{a}^\dagger\hat{a}\mathcal{L}_{\rm c}\rho(t)} +
    \Tr{\hat{H}_{\rm TD}'\mathcal{L}'\rho(t)} =J_{\rm c} + J' \fullstop
\end{equation}

We note that this decomposition analogously holds for the new framework, where we need to replace $\mathcal{L}_{\rm c}\rightarrow\mathcal{L}_{\rm s}$.

\section{III. Energy exchanged with the input and output fields}
We consider the sum of $P$ and $J_{\rm c}$, 
\begin{equation}
     P+J_{\rm c} = -\sqrt{\kappa}\omega_{\rm d} \bigg( \langle \hat{b}_{\text{in}}^\dagger(t) \rangle \langle \hat{a} \rangle + \langle \hat{b}_{\text{in}}(t) \rangle \langle \hat{a}^\dagger  \rangle \bigg)  + \omega_{\rm d}\kappa (n_{\rm c} - \langle \hat{a}^\dagger \hat{a} \rangle) \comma
\end{equation}
We may eliminate the cavity operator using $\sqrt{\kappa}\hat{a} = \bout - \bin$ (for ease of notation we suppress the time arguments) to find, 
\begin{align}
     P+J_{\rm c} &=\omega_{\rm d}\kappa n_{\rm c} - \omega_{\rm d} \bigg( \langle \hat{b}_{\text{in}}^\dagger \rangle \langle \bout \rangle - \langle \hat{b}_{\text{in}}^\dagger \rangle \langle \bin \rangle + \langle \hat{b}_{\text{in}} \rangle \langle \dbout  \rangle -  \langle \hat{b}_{\text{in}}(t) \rangle \langle \dbin  \rangle  \bigg)  \\ &~~- \omega_{\rm d}\bigg(\langle \dbout \bout \rangle -\langle \dbout \bin \rangle -\langle \dbin \bout \rangle +\langle \dbin \bin \rangle    \bigg)\nonumber\\
     &= \omega_{\rm d}\bigg( \kappa n_{\rm c} - \langle \dbout \bout \rangle + \langle \dbin \bin \rangle + \langle \!\langle  \dbout \bin  \rangle \!\rangle+ \langle\! \langle  \bout \dbin  \rangle\! \rangle - 2  \langle \!\langle \hat{b}_{\text{in}}  \dbin  \rangle \!\rangle \bigg) \nonumber 
\end{align}
Using the relation $\langle \!\langle  \dbin \hat{a} \rangle \!\rangle  = \langle \!\langle  \hat{a}^\dag \bin \rangle\! \rangle = -\sqrt{\kappa} \frac{n_{\rm c}}{2}$ \cite{gardiner_1985}, we find
\begin{equation}
\label{eq:inoutgard}
 \langle\! \langle \dbout \bin \rangle\! \rangle = \langle \!\langle \dbin \bout \rangle\! \rangle = \langle \!\langle \dbin \bin \rangle\! \rangle - \frac{{\kappa}}{2} n_{\rm c} \comma
\end{equation}
which results in
\begin{equation}
    P+J_{\rm c}  =\omega_{\rm d}\left(\langle\dbout(t)\bout(t)\rangle-\langle\dbin(t)\bin(t)\rangle\right)=P^{\rm io}+J^{\rm io}.
\end{equation}

Similarly, we may use $\sqrt{\kappa}\hat{a} = \bout - \bin$ and Eq.~\eqref{eq:inoutgard} to show
\begin{equation}
    J_{\rm c}^{\rm io} = -\omega_{\rm d}\left(\langle\!\langle\dbout(t)\bout(t)\rangle\!\rangle -\langle\!\langle\dbin(t)\bin(t)\rangle\!\rangle\right) = \omega_{\rm d}\kappa (n_{\rm c} - \langle\!\langle \hat{a}^\dagger \hat{a} \rangle\! \rangle) .
\end{equation}

\section{IV. Proof of the second law of thermodynamics}
To prove the second law of thermodynamics, we use Spohn's inequality \cite{spohn1978},
\begin{align}
    -\Tr{\mathcal{L}\rho(t) \qty[\ln\rho(t)-\ln \hat{\sigma}]} \geq 0 \comma \label{eq:spohn_ineq}
\end{align}
where $\mathcal{L}$ is the generator of a positive semi group and $\hat{\sigma}$ is its steady state, i.e., $\mathcal{L} \hat{\sigma} = 0$.

\subsection{Conventional framework}
We introduce thesteady states of the dissipators in the master equation in Eq.~(4)
\begin{equation}
    \mathcal{L}_{\rm c}\hat{\sigma}_{\rm c} = \mathcal{L}'\hat{\sigma}'=0,
\end{equation}
with
\begin{equation}
    \hat{\sigma}_{\rm c} =\frac{e^{-\omega_{\rm d}\hat{a}^\dagger\hat{a}/(k_{\rm B}T_{\rm c})}}{{\rm Tr}\left\{e^{-\omega_{\rm d}\hat{a}^\dagger\hat{a}/(k_{\rm B}T_{\rm c})}\right\}},\hspace{1.5cm}\hat{\sigma}' = \frac{e^{-\hat{H}_{\rm TD}'/(k_{\rm B}T')}}{{\rm Tr}\left\{e^{-\hat{H}_{\rm TD}'/(k_{\rm B}T')}\right\}}.
\end{equation}
While $\hat{\sigma}_{\rm c}$ is determined by our choice of $\mathcal{L}_{\rm c}$, c.f.~Eq.~(5), the stationary state $\hat{\sigma}'$ is assumed to be consistent with the thermodynamic Hamiltonian.

The heat currents $J_{\rm c}$ and $J'$ may consequently be written in terms of the dissipators and their stationary states
\begin{equation} 
    J_{\rm c} = - k_{\rm B} T_{\rm c} \Tr{ \mathcal{L}_{\rm c} \rho(t)\ln \hat{\sigma}_{\rm c}}\comma \label{eq:heatsecondlaw} \hspace{1.5cm}
    J'= - k_{\rm B} T'\Tr{ \mathcal{L}'\rho(t) \ln \hat{\sigma}'}    \fullstop 
\end{equation}
Substituting these heat currents into Eq.~(8) results in
\begin{align}
    \dot{\Sigma}/k_{\rm B} =  \partial_t S_{\text{vN}}(\rho(t)) + \Tr{ \mathcal{L}_{\rm c} \rho(t)\ln \hat{\sigma}_{\rm c}} + \Tr{ \mathcal{L}'\rho(t) \ln \hat{\sigma}'}     \comma 
\end{align}
which in turn can be written as
\begin{align}
    \dot{\Sigma}/k_{\rm B} &= -\Tr{\mathcal{L}\rho(t)\ln\rho(t)} + \Tr{ \mathcal{L}_{\rm c}\rho(t)\ln \hat{\sigma}_{\rm c}} + \Tr{ \mathcal{L}'\rho(t) \ln \hat{\sigma}'}  \nonumber \\
     &=-\Tr{\mathcal{L}_{\rm c}\rho(t) \qty[\ln\hat{\rho}(t)-\ln \hat{\sigma}_{\rm c}]}
     -\Tr{\mathcal{L}'\rho(t) \qty[\ln\hat{\rho}(t)-\ln \hat{\sigma}'\ ]} \geq 0
    \fullstop 
\end{align}
The result on the right hand side is a sum of two Spohn inequalities, see Eq.~\eqref{eq:spohn_ineq}. Therefore we have proven that the entropy production rate is indeed non-negative and the second law is verified for the conventional definitions of power and heat.

\subsection{New framework}
To derive the second law for the new framework we show that we can still write the new definition of the heat current as we do in equation \eqref{eq:heatsecondlaw}. To this end, we introduce the steady state of the shifted dissipator
\begin{equation}
   \mathcal{L}_{\rm s} \hat{\sigma}_{\rm s} =0,\hspace{1.5cm} \hat{\sigma}_{\rm s}=\frac{e^{-\omega_{\rm d} \left(\hat{a}^\dagger - \langle \hat{a}^\dagger \rangle )( \hat{a} - \langle \hat{a} \rangle \right)/\left(k_{\rm B}T_{\rm c}\right)}}{{\rm Tr}\left\{e^{-\omega_{\rm d} \left(\hat{a}^\dagger - \langle \hat{a}^\dagger \rangle )( \hat{a} - \langle \hat{a} \rangle \right)/\left(k_{\rm B}T_{\rm c}\right)}\right\}} \comma
\end{equation}
The new definition of the heat current is given by
\begin{equation}
    J_{\rm c}^{\rm io} = \Tr{\omega_{\rm d} \hat{a}^\dagger \hat{a} \mathcal{L}_{\rm s}\rho(t)} 
\end{equation}
We use the relations 
\begin{equation}
    \Tr{\hat{a} \mathcal{L}_{\rm s}\rho(t)} = \Tr{\hat{a}^\dagger \mathcal{L}_{\rm s}\rho(t)} = \Tr{ \mathcal{L}_{\rm s}\rho(t)} =0 \comma
\end{equation}
to show that the heat current can be written as 
\begin{equation}
    J_{\rm c}^{\rm io} =  \Tr{\omega_{\rm d} (\hat{a}^\dagger - \langle \hat{a}^\dagger \rangle )( \hat{a} - \langle \hat{a} \rangle)\mathcal{L}_{\rm s}\rho(t)} = - k_{\rm B} T_{\rm c} \Tr{ \mathcal{L}_{\rm s} \rho(t)\ln \hat{\sigma}_{\rm s}}\fullstop 
\end{equation}
The second law then follows analogously to how it is derived above for the conventional framework by writing
\begin{align}
   \dot{\Sigma}^{\rm io}\equiv k_{\rm B}\partial_tS_{\rm vN}(\rho(t))-\frac{J_{\rm c}^{\rm io}}{T_{\rm c}}-\frac{J'}{T'} = - k_{\rm B}\Tr{\mathcal{L}_{\rm s}\rho(t) \qty[\ln\hat{\rho}(t)- \ln \hat{\sigma}_{\rm s}]}
     - k_{\rm B}\Tr{\mathcal{L}'\rho(t) \qty[\ln\hat{\rho}(t)-\ln \hat{\sigma}'\ ]} \geq 0
    \fullstop 
\end{align}
We finally note that 
\begin{equation}
  \dot{\Sigma}-\dot{\Sigma}^{\rm io} =   \frac{J_{\rm c}^{\rm io}-J_{\rm c}}{T_{\rm c}} = \frac{\omega_d\kappa}{T_{\rm c}}|\langle \hat{a}\rangle|^2\geq 0,
\end{equation}

\section{V. Generalization of the framework}
\subsection{Master equation and thermodynamic Hamiltonian}
In this section, we extend the framework introduced in the main text to additional dissipation channels and multiple temperatures. To this end, we consider the master equation 
\begin{equation}
    \partial_t \hat{\rho}(t) = -i \qty[\hat{H}(t),\hat{\rho}(t)] + \mathcal{L}\hat{\rho}(t)(t)\comma\label{eq:system_QMEapp}
\end{equation}
with the dissipator
\begin{equation}
    \label{eq:dissgen}
    \mathcal{L} = \sum_j\mathcal{L}_{{\rm c},j}+\sum_k\mathcal{L}_{k}+\sum_l\mathcal{L}_l'.
\end{equation}
Here 
\begin{align}
    \mathcal{L}_{{\rm c},j} \hat{\rho} &= \kappa_j n_{{\rm c},j} \mathcal{D}[\hat{a}^\dagger] \hat{\rho} + \kappa_j (n_{{\rm c},j}+1) \mathcal{D}[\hat{a}]\hat{\rho}  \comma 
\end{align}
denote photon loss due to output channels that are accessible. The coherent field leaking out into these channels are therefore interpreted as work. In addition, we consider inaccessible channels where photons are exchanged with the environment described by the dissipators
\begin{align}
    \mathcal{L}_{k} \hat{\rho} &= \gamma_k n_{k} \mathcal{D}[\hat{a}^\dagger] \hat{\rho} + \gamma_k(n_{k}+1) \mathcal{D}[\hat{a}]\hat{\rho}.
\end{align}
The dissipators $\mathcal{L}'_l$ denote dissipation of the system embedded into the cavity. Splitting this dissipator into multiple parts allows us to consider reservoirs at different temperatures.

The Hamiltonian in Eq.~\eqref{eq:system_QMEapp} reads
\begin{equation}
    \label{eq:hamgen}
    \hat{H}(t) = \Omega\hat{a}^\dagger\hat{a}+\hat{H}'(t)+i\sum_j\sqrt{\kappa_j}\left(\langle \hat{b}^\dagger_{{\rm in},j}(t)\rangle\hat{a} -\langle \hat{b}_{{\rm in},j}(t)\rangle\hat{a}^\dagger \right)+i\sum_k\sqrt{\gamma_k}\left(\langle \hat{c}^\dagger_{{\rm in},k}(t)\rangle\hat{a} -\langle \hat{c}_{{\rm in},k}(t)\rangle\hat{a}^\dagger \right),
\end{equation}
where $\hat{b}_{{\rm in},j}$ and $\hat{c}_{{\rm in},j}$ denote the input modes of the observed and unobserved channels respectively. Note that we also allow $\hat{H}'(t)$ to be explicitly time-dependent. As in the main text, we leave $\hat{H}'(t)$ and $\mathcal{L}_l'$ unspecified but we assume that a thermodynamic description based on a single thermodynamic Hamiltonian exists. To this end, we make the following assumptions
\begin{equation}
    {\rm Tr}\left\{\hat{a}^\dagger\hat{a}\mathcal{L}'_l\hat{\rho}\right\} = {\rm Tr}\left\{\hat{a}^{(\dagger)}\mathcal{L}'_l\hat{\rho}\right\}=0.
\end{equation}

We use the same thermodynamic Hamiltonian as in the main text
\begin{align}
\hat{H}_{\rm TD} = \omega_{\rm d} \hat{a}^\dagger \hat{a} + \hat{H}_{\rm TD}'  \comma \label{eq:TD_hamiltonianapp}
\end{align}
where we assume that
\begin{equation}
    [\hat{H}_{\rm TD}',\hat{a}] = 0,\hspace{2cm}\mathcal{L}_l'\hat{\sigma}'_l = 0, \hspace{.5cm}\hat{\sigma}'_l =\frac{e^{-\hat{H}_{\rm TD}'/(k_{\rm B}T'_l)}}{{\rm Tr}\left\{e^{-\hat{H}_{\rm TD}'/(k_{\rm B}T'_l)}\right\}}.
\end{equation}
In addition, one may show that
\begin{equation}
   \mathcal{L}_{{\rm c},j}\hat{\sigma}_{{\rm c},j} = 0,\hspace{.5cm}  \mathcal{L}_{k}\hat{\sigma}_{k} = 0,\hspace{.5cm}\hat{\sigma}_{{\rm c},j} =\frac{e^{-\omega_{\rm d}\hat{a}^\dagger\hat{a}/(k_{\rm B}T_{{\rm c},j})}}{{\rm Tr}\left\{e^{-\omega_{\rm d}\hat{a}^\dagger\hat{a}/(k_{\rm B}T_{{\rm c},j})}\right\}},\hspace{.5cm}\hat{\sigma}_{k} =\frac{e^{-\omega_{\rm d}\hat{a}^\dagger\hat{a}/(k_{\rm B}T_{k})}}{{\rm Tr}\left\{e^{-\omega_{\rm d}\hat{a}^\dagger\hat{a}/(k_{\rm B}T_{k})}\right\}},
\end{equation}
where
\begin{equation}
n_{{\rm c},j} =\frac{1}{e^{\omega_d/(k_{\rm B}T_{{\rm c},j})}-1},\hspace{1cm}n_{k} =\frac{1}{e^{\omega_d/(k_{\rm B}T_{k})}-1}.
\end{equation}
We further assume that
\begin{equation}
    -i[\hat{H}_{\rm TD},\hat{H}(t)]\simeq\partial_t\hat{H}(t).
\end{equation}
This is the case as long as $-i[\hat{H}_{\rm TD},\hat{H}'(t)]\simeq\partial_t\hat{H}'(t)$ and all input drives only contain frequencies close to $\omega_{\rm d}$. If the input drives contain frequencies far from $\omega_{\rm d}$, we may always drop them as they will not be able to enter the cavity in the first place. Indeed, we may choose $\omega_{\rm d}=\Omega$, in particular if there is no unique drive frequency.

\subsection{Conventional framework}
Like in the main text, the first law reads
\begin{equation}
    \partial_t\langle \hat{H}_{\rm TD}\rangle = P+J,\hspace{1.5cm}P = -i\langle[\hat{H}_{\rm TD},\hat{H}(t)]\rangle,\hspace{.5cm}J = {\rm Tr}\left\{\hat{H}_{\rm TD}\mathcal{L}\hat{\rho}(t)\right\}.
\end{equation}
With the assumptions above, we may divide the power into different contributions
\begin{equation}
    P = \sum_j P_{{\rm c},j}+\sum_k P_k+P',
\end{equation}
with
\begin{equation}
    P_{{\rm c},j}= -\sqrt{\kappa_j}\omega_{\rm d} \bigg( \langle \hat{b}_{\text{in},j}^\dagger(t) \rangle \langle \hat{a} \rangle + \langle \hat{b}_{\text{in},j}(t) \rangle \langle \hat{a}^\dagger  \rangle \bigg),\hspace{.5cm}P_k= -\sqrt{\gamma_k}\omega_{\rm d} \bigg( \langle \hat{c}_k^\dagger(t) \rangle \langle \hat{a} \rangle + \langle \hat{c}_k(t) \rangle \langle \hat{a}^\dagger  \rangle \bigg),
\end{equation}
and
\begin{equation}
    P' = -i\langle[\hat{H}_{\rm TD}',\hat{H}'(t)]\rangle.
\end{equation}
Similarly, heat can be divided as
\begin{equation}
    J = \sum_j J_{{\rm c},j}+\sum_k J_k+\sum_l J'_l,
\end{equation}
with
\begin{equation}
    J_{{\rm c},j} = \omega_{\rm d}\kappa_j (n_{{\rm c},j} - \langle \hat{a}^\dagger \hat{a} \rangle),\hspace{1.5cm} J_k = \omega_{\rm d}\gamma_k (n_k - \langle \hat{a}^\dagger \hat{a} \rangle),\hspace{1.5cm}J'_l = {\rm Tr}\left\{\hat{H}'_{\rm TD}\mathcal{L}'_l\hat{\rho}(t)\right\}.
\end{equation}

The second law of thermodynamics may be written as
\begin{equation}
    \label{eq:entprodgen}
    \dot{\Sigma} \equiv k_{\rm B}\partial_tS_{\rm vN}(\hat{\rho})-\sum_j\frac{J_{{\rm c},j}}{T_{{\rm c},j}}-\sum_k\frac{J_{k}}{T_{k}}-\sum_l\frac{J'_l}{T'_l} \geq 0 .
\end{equation}
It can be proven by casting the entropy production into
\begin{equation}
     \dot{\Sigma}/{k_{\rm B}} = -\sum_j\Tr{\mathcal{L}_{{\rm c},j}\rho(t) \qty[\ln\hat{\rho}(t)-\ln \hat{\sigma}_{{\rm c},j}]}-\sum_k\Tr{\mathcal{L}_{k}\rho(t) \qty[\ln\hat{\rho}(t)-\ln \hat{\sigma}_{k}]}
     -\sum_l\Tr{\mathcal{L}'_l\rho(t) \qty[\ln\hat{\rho}(t)-\ln \hat{\sigma}'_l\ ]},
\end{equation}
and using Spohn's inequality.

\subsection{New framework}
In the new framework, the first law reads
\begin{equation}
    \partial_t\langle \hat{H}_{\rm TD}\rangle = \sum_{j}\left(P_{{\rm c},j}^{\rm io} +J_{{\rm c},j}^{\rm io}\right)+\sum_k\left(P_k+J_k\right)+P'+\sum_lJ_l',
\end{equation}
where
\begin{equation}
     P_{{\rm c},j}^{\rm io} =- \omega_{\rm d}\left(|\langle\hat{b}_{{\rm out},j}(t)\rangle|^2 -|\langle\hat{b}_{{\rm in},j}(t)\rangle|^2\right), \hspace{1.5cm} J_{{\rm c},j}^{\rm io} = -\omega_{\rm d}\left(\langle\!\langle\hat{b}_{{\rm out},j}^\dagger(t)\hat{b}_{{\rm out},j}(t)\rangle\!\rangle -\langle\!\langle\hat{b}_{{\rm in},j}^\dagger(t)\hat{b}_{{\rm in},j}(t)\rangle\!\rangle\right),
\end{equation}
and the other terms remain the same as in the conventional framework. Note that we only altered the terms that correspond to output channels that we assume are accessible. In a straightforward generalization of the derivation given in Sec.~III above, we find
\begin{equation}
    P_{{\rm c},j}^{\rm io}+J_{{\rm c},j}^{\rm io}=P_{{\rm c},j}+J_{{\rm c},j},
\end{equation}
and
\begin{equation}
    J_{{\rm c},j}^{\rm io} = \omega_{\rm d}\kappa_j (n_{{\rm c},j} - \langle\!\langle \hat{a}^\dagger \hat{a} \rangle\!\rangle).
\end{equation}

As in the main text, we obtain the laws of thermodyanmics from re-writing the master equation as
\begin{equation}
    \partial_t \hat{\rho}(t) = -i\left[\hat{H}_{\rm s}(t),\hat{\rho}(t)\right] +\left[\sum_j\mathcal{L}_{{\rm s},j}+\sum_k\mathcal{L}_{k}+\sum_l\mathcal{L}_l'\right]\hat{\rho}(t),
\end{equation}
with
\begin{equation}
    \hat{H}_{\rm s}(t) = \hat{H} +\frac{i}{2}\sum_j\kappa_j\left(\langle \hat{a}^\dagger\rangle \hat{a}  - \langle \hat{a}\rangle \hat{a}^\dagger \right),
\end{equation}
and
\begin{equation}
   \mathcal{L}_{{\rm s},j}  = \kappa_j n_{{\rm c},j} \mathcal{D}[\hat{a}^\dagger - \langle \hat{a}^\dagger \rangle] 
    + \kappa_j (n_{{\rm c},j}+1) \mathcal{D}[\hat{a} - \langle \hat{a} \rangle].
\end{equation}
The steady state of this dissipator reads
\begin{equation}
   \mathcal{L}_{{\rm s},j} \sigma_{\rm s} =0,\hspace{1.5cm} \sigma_{{\rm s},j}=\frac{e^{-\omega_{\rm d} \left(\hat{a}^\dagger - \langle \hat{a}^\dagger \rangle )( \hat{a} - \langle \hat{a} \rangle \right)/\left(k_{\rm B}T_{{\rm c},j}\right)}}{{\rm Tr}\left\{e^{-\omega_{\rm d} \left(\hat{a}^\dagger - \langle \hat{a}^\dagger \rangle )( \hat{a} - \langle \hat{a} \rangle \right)/\left(k_{\rm B}T_{{\rm c},j}\right)}\right\}}.
\end{equation}

The power and the contributions to the heat current may then be obtained through
\begin{equation}
    -i\langle[\hat{H}_{\rm TD},\hat{H}_{\rm s}(t)]\rangle = \sum_{j}P_{{\rm c},j}^{\rm io} +\sum_kP_k,\hspace{1.5cm} J_{{\rm c},j}^{\rm io} = {\rm Tr}\left\{\hat{H}_{\rm TD}\mathcal{L}_{{\rm s},j}\hat{\rho}(t)\right\}.
\end{equation}
The entropy production reads
\begin{equation}
    \label{eq:entprodgenio}
    \dot{\Sigma}^{\rm io} \equiv k_{\rm B}\partial_tS_{\rm vN}(\hat{\rho})-\sum_j\frac{J_{{\rm c},j}^{\rm io}}{T_{{\rm c},j}}-\sum_k\frac{J_{k}}{T_{k}}-\sum_l\frac{J'_l}{T'_l} \geq 0 .
\end{equation}
And the second law is again proven using Spohn's inequality by casting the entropy production into
\begin{equation}
     \dot{\Sigma}/{k_{\rm B}} = -\sum_j\Tr{\mathcal{L}_{{\rm s},j}\rho(t) \qty[\ln\hat{\rho}(t)-\ln \hat{\sigma}_{{\rm c},j}]}-\sum_k\Tr{\mathcal{L}_{k}\rho(t) \qty[\ln\hat{\rho}(t)-\ln \hat{\sigma}_{k}]}
     -\sum_l\Tr{\mathcal{L}'_l\rho(t) \qty[\ln\hat{\rho}(t)-\ln \hat{\sigma}'_l\ ]}.
\end{equation}
To this end, we used
\begin{equation}
    \Tr{\omega_{\rm d} \hat{a}^\dagger \hat{a} \mathcal{L}_{{\rm s},j}\rho(t)} = \Tr{\omega_{\rm d} (\hat{a}^\dagger - \langle \hat{a}^\dagger \rangle )( \hat{a} - \langle \hat{a} \rangle)\mathcal{L}_{{\rm s},j}\rho(t)}.
\end{equation}
We further find
\begin{equation}
  \dot{\Sigma}-\dot{\Sigma}^{\rm io} =   \sum_j\frac{J_{{\rm c},j}^{\rm io}-J_{{\rm c},j}}{T_{{\rm c},j}} = \omega_d\sum_j\frac{\kappa_j}{T_{{\rm c},j}}|\langle \hat{a}\rangle|^2\geq 0,
\end{equation}

\section{VI. Entropy production as loss of information}
Here we go beyond a Markovian description, starting with the Hamiltonian that describes system and environment
\begin{equation}
    \hat{H}_{\rm tot} = \hat{H} + \hat{H}_{\rm B} + \hat{V}+H_{\rm B}'+V', \label{eq:hamtot}
\end{equation}
where $\hat{H}$ is the Hamiltonian of the cavity, together with anything embedded in it. For simplicity, we consider a single bosonic bath coupled to the cavity
\begin{equation}
    \hat{H}_{\rm B} = \sum_k \omega_k \hat{b}_k^\dagger \hat{b}_k, \hspace{1.5cm} \hat{V} = \sum_k \qty( g_k \hat{a}^\dagger \hat{b}_k + g_k^* \hat{b}_k^{\dagger} \hat{a}),
    \label{eq:general_setting_bath}
\end{equation}
generalizing to multiple baths is straightforward. The environment of the intra-cavity system is described by the Hamiltonian $\hat{H}_{\rm B}'$ and coupling $\hat{V}'$.

In the conventional framework, the entropy production may be defined as~\cite{Esposito_2010}
\begin{equation}
    \label{eq:entcon}
    \Sigma(t) = k_BS[\hat{\rho}_{\rm tot}(t)||\hat{\rho}(t)\otimes\hat{\tau}]\geq0,
\end{equation}
where $S[\hat{\rho}||\hat{\sigma}]={\rm Tr}\{\hat{\rho}\ln\hat{\rho}-\hat{\rho}\ln\hat{\sigma}\}$ is the quantum relative entropy and we introduced the Gibbs state of the bath
\begin{equation}
    \hat{\tau}_ = \frac{e^{-\beta_{\rm c}\hat{H}_{\rm B}-\beta'\hat{H}_{\rm B}'}}{{\rm Tr}\left\{e^{-\beta_{\rm c}\hat{H}_{\rm B}-\beta'\hat{H}_{\rm B}'}\right\}}.
\end{equation}
In this framework, the cavity Hamiltonian includes the drive
\begin{equation}
    \hat{H}(t) = \Omega\hat{a}^\dagger\hat{a}+i\sqrt{\kappa} \left(\langle \dbin(t)\rangle\hat{a} -\langle\bin(t)\rangle\hat{a}^\dagger \right)+\hat{H}',
\end{equation}
and the total density matrix reads
\begin{equation}
\label{eq:totunitary}
    \hat{\rho}_{\rm tot}(t) = \hat{U}(t)\hat{\rho}_{\rm tot}(0)\hat{U}^\dagger(t),\hspace{2cm}\hat{U}(t)=\mathcal{T}e^{-i\int_{0}^tdt'\hat{H}_{\rm tot}(t')},
\end{equation}
where $\mathcal{T}$ denotes time-ordering. The reduced density matrix in Eq.~\eqref{eq:entcon} is then obtained through $\hat{\rho}(t)={\rm Tr}_{\rm B}\{\hat{\rho}_{\rm tot}(t)\}$. To ensure that $\Sigma(0)=0$, we assume an initial state of the form $\hat{\rho}_{\rm tot}(0)=\hat{\rho}(0)\otimes\hat{\tau}$. The nonnegativity of the entropy production is a direct consequence of the nonnegativity of the quantum relative entropy. The entropy production in Eq.~\eqref{eq:entcon} can be understood as the information that is lost when using an effective description provided by $\hat{\rho}(t)\otimes\hat{\tau}$~\cite{potts2024quantumthermodynamics}, i.e., where we describe the system by its exact reduced density matrix but the environment by a Gibbs state that is fully determined by the initial temperature. From Eq.~\eqref{eq:entcon}, we can derive
\begin{equation}
    \dot{\Sigma} = k_B\partial_tS_{\rm vN}(\hat{\rho})-\frac{J_{\rm c}}{T_{\rm c}}-\frac{J'}{T'},
\end{equation}
where the heat currents are defined as
\begin{equation}
    \label{eq:heatsunitary}
    J_{\rm c} = -\partial_t{\rm Tr}\{\hat{H}_{\rm B}\hat{\rho}_{\rm tot}(t)\},\hspace{2cm}J' = -\partial_t{\rm Tr}\{\hat{H}'_{\rm B}\hat{\rho}_{\rm tot}(t)\}.
\end{equation}
Performing the approximations that result in the Markovian master equation then results in the definitions of heat in the main text, see Ref.~\cite{Potts_2021}.

As we now show, we can obtain our new thermodynamic framework in an analogous way, where entropy production is defined as the quantum relative entropy between the exact density matrix and an effective description
\begin{equation}
    \label{eq:entiocon}
    \Sigma^{\rm io}(t) = k_BS[\hat{\rho}_{\rm tot}(t)||\hat{\rho}(t)\otimes\hat{\tau}({\mathbf \alpha)}]\geq 0,
\end{equation}
where the effective description of the environment is provided by a displaced coherent state
\begin{equation}
\label{eq:taualpha}
    \hat{\tau}(\alpha) = \frac{e^{-\beta_{\rm c}\sum_k\omega_k\left(\hat{b}^\dagger_k-\alpha^*_k\right)\left(\hat{b}_k-\alpha_k\right)}}{{\rm Tr}\left\{e^{-\beta_{\rm c}\sum_k\omega_k\left(\hat{b}^\dagger_k-\alpha^*_k\right)\left(\hat{b}_k-\alpha_k\right)}\right\}}\frac{e^{-\beta'\hat{H}'_{\rm B}}}{{\rm Tr}\left\{e^{-\beta'\hat{H}'_{\rm B}}\right\}}.
\end{equation}
Here the displacement is chosen such that it reflects the exact displacement, i.e.,
\begin{equation}
    \alpha_k = {\rm Tr}\{\hat{b}_k\hat{\rho}_{\rm tot}(t)\},
\end{equation}
and is thus a time-dependent quantity. In this framework, the cavity Hamiltonian does not include the coherent drive
\begin{equation}
\label{eq:hamsys}
    \hat{H} = \Omega\hat{a}^\dagger\hat{a}+\hat{H}',
\end{equation}
which is instead provided by the initial displacement of the environment. For consistency, and to ensure that $\Sigma(0)=0$, we choose an initial state that matches the effective description, i.e. $\hat{\rho}_{\rm tot}(0)=\hat{\rho}(0)\otimes\hat{\tau}(\alpha)$. Time evolution of the total density matrix is again given by unitary evolution with the total Hamiltonian, see Eq.~\eqref{eq:totunitary}. With the help of Eq.~\eqref{eq:taualpha}, we find
\begin{equation}
    \label{eq:sigdotgen}
    \dot{\Sigma}^{\rm io} = k_B\partial_tS_{\rm vN}(\hat{\rho})-\frac{J_{\rm c}^{\rm io}}{T_{\rm c}}-\frac{J'}{T'},
\end{equation}
where $J'$ is given by Eq~\eqref{eq:heatsunitary} and the cavity heat current reads
\begin{equation}
    \label{eq:cavityheatunit}
    J_{\rm c}^{\rm io} = -\partial_t\sum_{k}\omega_k\langle\!\langle \hat{b}_{k}^\dagger\hat{b}_{k}\rangle\!\rangle,
\end{equation}
where the covariance is taken with respect to the (exact) total density matrix, i.e., $\langle\bullet\rangle = {\rm Tr}\{\bullet\hat{\rho}_{\rm tot}(t)\}$. To show that this heat current is indeed the same as in the main text under the Markovian approximations, we first note that only modes with frequencies close to $\omega_{\rm d}$ (which is close to the cavity frequency) will be affected. We may therefore write
\begin{equation}
    \label{eq:cavityheatunit2}
    J_{\rm c}^{\rm io} \simeq -\omega_{\rm d}\partial_t\sum_{k}\langle\!\langle \hat{b}_{k}^\dagger\hat{b}_{k}\rangle\!\rangle.
\end{equation}
To evaluate this expression, we first consider
\begin{equation}
    \partial_t\langle\hat{b}_k\rangle = -i\omega_k\langle\hat{b}_k\rangle-ig_k^*\langle\hat{a}\rangle = -i\omega_k\langle\hat{b}_k\rangle-\frac{ig_k^*}{\sqrt{\kappa}}\left(\langle\hat{b}_{\rm out}\rangle-\langle\hat{b}_{\rm in}\rangle\right),
\end{equation}
where we used unitary dynamics under the Hamiltonian given by Eqs.~\eqref{eq:hamtot} and \eqref{eq:hamsys}, as well as the input-output relation $\hat{b}_{\rm out}=\hat{b}_{\rm in}+\sqrt{\kappa}\hat{a}$. Similarly, we find
\begin{equation}
    \partial_t\langle\hat{b}_k^\dagger\hat{b}_k\rangle =ig_k\langle\hat{a}^\dagger\hat{b}_k\rangle - ig_k^*\langle\hat{b}_k^\dagger\hat{a}\rangle.
\end{equation}
To make progress, we solve the Heisenberg equation for the bath modes under unitary evolution which results in
\begin{equation}
    \hat{b}_k(t) = e^{-i\omega_k(t-t_0)}\hat{b}_k(t_0)-ig_k^*\int_{t_0}^td\tau e^{-i\omega_k(t-\tau)}\hat{a}(\tau),
\end{equation}
where $t_0$ is a time in the distant past. With the help of this solution, we find
\begin{equation}
\label{eq:igk}
    \sum_kig_k\hat{b}_k(t) = \sqrt{\kappa}\hat{b}_ {\rm in}(t)+\int_{t_0}^td\tau\int d\omega e^{-i\omega(t-\tau)}\rho(\omega)\hat{a}(\tau)\simeq \sqrt{\kappa}\hat{b}_ {\rm in}(t)+\frac{\kappa}{2}\hat{a}(t).
\end{equation}
Here we introduced the input operator in terms of the bath modes as
\begin{equation}
    \hat{b}_{\rm in}(t) = \frac{1}{\sqrt{\kappa}}\sum_kig_ke^{-i\omega_k(t-t_0)}\hat{b}_k(t_0),
\end{equation}
as well as the spectral density
\begin{equation}
    \rho(\omega)=\sum_k|g_k|^2\delta(\omega-\omega_k)\simeq\frac{\kappa}{2\pi},
\end{equation}
where in the last approximation, we assume that we can replace the spectral density with a constant under the integral. With the help of Eq.~\eqref{eq:igk} and the input-output relation, we find
\begin{equation}
    \partial_t\sum_k\langle\hat{b}_k^\dagger\hat{b}_k\rangle = \langle\hat{b}^\dagger_{\rm out}\hat{b}_{\rm out}\rangle-\langle\hat{b}^\dagger_{\rm in}\hat{b}_{\rm in}\rangle,
\end{equation}
and
\begin{equation}
    \partial_t\sum_k\langle \hat{b}_k^\dagger\rangle\langle\hat{b}_k\rangle = |\langle\hat{b}_{\rm out}\rangle|^2-|\langle\hat{b}_{\rm in}\rangle|^2.
\end{equation}
Together these equations imply
\begin{equation}
    J_{\rm c}^{\rm io} = -\omega_{\rm d}\partial_t\sum_{k}\langle\!\langle \hat{b}_{k}^\dagger\hat{b}_{k}\rangle\!\rangle=  -\omega_{\rm d}\left(\langle\!\langle\hat{b}^\dagger_{\rm out}\hat{b}_{\rm out}\rangle\!\rangle-\langle\!\langle\hat{b}^\dagger_{\rm in}\hat{b}_{\rm in}\rangle\!\rangle\right),
\end{equation}
recovering the expression from the main text.

\end{document}